\documentclass{aastex631}

\shorttitle{IXPE view of H 1426+428}
\shortauthors{Banerjee et al.}
\graphicspath{{./}{figures/}}
\begin{document}

\title{Contemporaneous X-ray and Optical Polarization of EHSP Blazar H 1426+428}

\author[0000-0001-7796-8907]{Anuvab Banerjee}
\affiliation{Clemson University, Clemson, SC 29634, USA}

\author{Akash Garg}
\affiliation{Inter-University Center for Astronomy and Astrophysics, Ganeshkhind, Pune 411007, India}

%\collaboration{6}{(AAS Journals Data Editors)}

\author[0000-0002-9280-2785]{Divya Rawat}
%\affiliation{Inter-University Center for Astronomy and Astrophysics, Ganeshkhind, Pune 411007, India}
\affiliation{Observatoire Astronomique de Strasbourg, Universite'de Strasbourg, CNRS, 11 rue de l'Universite, F-67000 Strasbourg, France}

\author{Svetlana Jorstad}
\affiliation{Institute for Astrophysical Research, Boston University, 725 Commonwealth Avenue, Boston, MA 02215, USA}
\affiliation{Saint Petersburg State University, 7/9 Universitetskaya nab., 199034 St. Petersburg, Russia.}

\author{Alan P. Marscher}
\affiliation{Institute for Astrophysical Research, Boston University, 725 Commonwealth Avenue, Boston, MA 02215, USA}

\author{Ivan Agudo}
\affiliation{Instituto de Astrof\'{i}sica de Andaluc\'{i}a (CSIC), Glorieta de la Astronom\'{i}a s/n, 18008 Granada, Spain\\}

\author{Jorge Otero-Santos}
\affiliation{Instituto de Astrof\'{i}sica de Andaluc\'{i}a (CSIC), Glorieta de la Astronom\'{i}a s/n, 18008 Granada, Spain\\}
\affiliation{Istituto Nazionale di Fisica Nucleare, Sezione di Padova, 35131 Padova, Italy}

\author{Daniel Morcuende}
\affiliation{Instituto de Astrof\'{i}sica de Andaluc\'{i}a (CSIC), Glorieta de la Astronom\'{i}a s/n, 18008 Granada, Spain\\}

\author{Juan Escudero Pedrosa}
\affiliation{Instituto de Astrof\'{i}sica de Andaluc\'{i}a (CSIC), Glorieta de la Astronom\'{i}a s/n, 18008 Granada, Spain\\}
\affiliation{Center for Astrophysics, Harvard \& Smithsonian, Cambridge, MA 02138, USA}

\author[0000-0002-3433-4610]{Alberto Dom\'inguez}
\affiliation{IPARCOS and Department of EMFTEL, Universidad Complutense de Madrid, E-28040 Madrid, Spain}

\author[0000-0002-2878-4025]{Ayan Bhattacharjee}
\affiliation{Research Institute of Basic Sciences, Seoul National University, Seoul 08826, Republic of Korea}

\author{Isaiah Cox}
\affiliation{Clemson University, Clemson, SC 29634, USA}

\author[0000-0002-7825-1526]{Indrani Pal}
\affiliation{Clemson University, Clemson, SC 29634, USA}

\author[0000-0002-7791-3671]{Xiurui Zhao}
\affiliation{Department of Astronomy, University of Illinois at Urbana-Champaign, Urbana, IL 61801, USA}
\affiliation{Cahill Center for Astrophysics, California Institute of Technology, 1216 East California Boulevard, Pasadena, CA 91125, USA}

\author[0000-0001-6412-2312]{Andrealuna Pizzetti}
\affiliation{European Southern Observatory, Alonso de C\'ordova 3107, Casilla 19, Santiago 19001, Chile}

\author[0000-0001-5544-0749]{Stefano Marchesi}
\affiliation{Dipartimento di Fisica e Astronomia (DIFA), Università di Bologna, via Gobetti 93/2, I-40129 Bologna, Italy}
\affiliation{Clemson University, Clemson, SC 29634, USA}
\affiliation{INAF, Osservatorio di Astrofisica e Scienza dello Spazio di Bologna, via P. Gobetti 93/3, 40129 Bologna, Italy}

\author{N\'uria Torres-Alb\`a}
\affiliation{Department of Astronomy, University of Virginia, P.O. Box 400325, Charlottesville, VA 22904, USA}

\author[0009-0003-3381-211X]{Kouser Imam}
\affiliation{Clemson University, Clemson, SC 29634, USA}

\author{Ross Silver}
\affiliation{NASA Goddard Space Flight Center, Greenbelt, MD 20771, USA}

\author[0000-0002-6584-1703]{Marco Ajello}
\affiliation{Clemson University, Clemson, SC 29634, USA}

%\author{others}
%\affiliation{to be added}


\begin{abstract}
We present the first contemporaneous X-ray and optical polarimetric measurement of the extremely high synchrotron peaked (EHSP) blazar H 1426+428. The X-ray polarimetric observations were undertaken using the Imaging X-ray Polarimetry Explorer (\textit{IXPE}) on 2024 May 27, and 2024 July 5. The \textit{IXPE} pointings were accompanied by contemporaneous optical observations of the Observatorio de Sierra Nevada, Calar Alto Observatory and the Perkins Telescope Observatory. While we observed the X-ray degree of polarization to be $>20\%$, the polarization in the optical band was found to be only $1-3\%$. This trend has been observed in several HSP blazars with available optical and X-ray polarimetric data and is typically explained in terms of energy stratification downstream of a shock. However, we observed a significant difference between the optical and X-ray polarization angles, a feature that
has been observed in certain HSP blazars, such as Mrk 421, but remains a relatively rare or underreported phenomenon. We discuss possible scenarios for these findings within the framework of a partially turbulent jet model. 

\end{abstract}



\section{Introduction} \label{sec:intro}
Blazars are a subclass of Active Galactic Nuclei (AGN) whose powerful jets are oriented towards our line of sight \citep[e.g.][]{urry1995unified}. The electromagnetic emission from such objects spans the entire electromagnetic spectrum, ranging from lower-energy radio to very high-energy $\gamma-$rays \citep{hovatta2019relativistic}. The broadband spectral energy distribution (SED) of blazars is characterized by the double-hump shape, with the low-energy synchrotron emission peaking in infra-red (IR)/X-ray regime, and the higher energy peak in the MeV-GeV regime \citep{fossati1998unifying}. High-synchrotron-peaked blazars form a sub-class with synchrotron peaks occurring at frequencies $\ge 10^{15}$ Hz, i.e. in the UV/X-ray regime \citep{abdo2010spectral}.  An HSP blazar with synchrotron peak frequency occurring at frequencies $\nu_{syn} \sim 10^{17}$ Hz is termed an extreme HSP (EHSP) blazar \citep{giommi1999sedentary,paliya2019fermi,nievas2022hunting}. \par 

With the launch of the Imaging X-ray Polarimetry Explorer \citep[IXPE,][]{weisskopf2022imaging}, comprehensive broadband polarization measurements of blazars from radio to X-rays have become
possible, which has enabled the community to infer the particle acceleration properties in blazar jets \citep{liodakis2022polarized,di2022x,di2023discovery}. Since the X-ray emission is associated with the high energy electrons close to the acceleration site, such measurements are extremely crucial \citep{chen2024x}. The first of such measurements on the well-known blazar source Mrk 501 revealed the degree of polarization (PD hereafter) in X-ray to be $\sim10\%$, higher compared to the optical polarization ($\sim4-5\%$), with the polarization angle (PA hereafter) oriented parallel to the jet axis \citep{liodakis2022polarized}. The result can be interpreted in terms of the synchrotron emission arising out of diffusive shock acceleration of an energy-stratified electron population. As the high-energy electrons propagate away from the acceleration site, they sample more turbulent segments of the jet, and consequently, the optical and radio emission becomes less polarized. Such a trend is observed in several other HSP blazars, including Mrk 421 \citep{di2022x}. \textbf{However, in general, there does not seem to be an obvious relationship between the polarization angle and the radio jet position angle. A mismatch between PA and the jet orientation has been detected in some blazar sources \citep[e.g. 1ES 0229+200;][]{ehlert2023x}, while evidence of time variability of PA has been observed in Mrk 421 \citep{di2023discovery}.}  \par 

H 1426+428 is categorized as an EHSP source with its synchrotron peak lying in the X-ray region \citep[$\nu_{syn} \sim 10^{18}$ Hz,][]{chang20193hsp}. The source is bright enough in the optical \citep[R-mag = 14.4,][]{veron2010catalogue} to allow for optical polarimetric observations from ground-based optical facilities. The source is also bright in X-ray with a flux  $\geq2.3\times10^{-11}~{ergs}~ cm^{-2}s^{-1}$ in the 2-8 keV band which allows polarimetric measurements using \textit{IXPE}. In addition, H 1426+428 is a TeV-detected HSP blazar \citep{aharonian2002tev,quinn2021veritas}. At a redshift of z = 0.129 \citep{remillard1989two}, H
1426+428 is relatively close, allowing for detailed studies of its
broadband emission without excessive extragalactic background light
attenuation \citep[e.g.][]{saldana2021observational,dominguez2024new}. Its proximity helps in studying environments of nearby galaxies with active blazars, which could help contextualize the evolution of similar sources at higher redshifts. For this source, particle acceleration by shock and magnetic reconnection has been proposed as an operative mechanism based on the intra-day variability signatures in X-ray \citep{devanand2022x} and optical \citep{chang2024optical} domains. Therefore, a comparison between optical and X-ray polarimetric signatures can break the degeneracy regarding the operative acceleration mechanism for this source. \par 

In this Letter, we report the results of the first \textit{IXPE} monitoring observation of H 1426+428. In Section 2, we provide the details of observation. In Section 3, we present the results, and in Section 4, we draw our conclusions.

\section{Observations and Data Analysis}
For the purpose of our analysis, we have used contemporaneous \textit{Swift}-XRT and \textit{IXPE} data. The analysis has been undertaken with using the FTOOLS software package HeaSoft version HEADAS-6.32 and XSPEC \citep{arnaud1996} version 12.13.1, with the latest calibration database (CALDB). The details of the observation for each instrument are provided in Table \ref{tab:obs}. Below, we describe the reduction and analysis techniques used in this work.

\subsection{IXPE}
\textit{IXPE}\footnote{\textit{IXPE} is capable of measuring the polarization properties, as well as spectral and temporal properties of X-ray sources in the 2-8 keV band \citep{weisskopf2022imaging}} has observed H 1426+428 on 2024 May 27, and 2024 July 5, for a total exposure of $\sim$210 ks (proposal no: 1186, PI: A. Banerjee). For the scientific analysis, we use cleaned and calibrated \textsc{level-2} event files for all the detector units (DU) obtained from the High Energy Astrophysics Science Archive Research Center (HEASARC) archive. In order to choose the appropriate source and background regions, we extract the imaging data from the event files using the \textsc{ftools} task \textsc{xselect}, and plot using the plotting software \textsc{ds9}. The IXPE \textsc{level-2} photon event list has been processed through the rejection algorithm of \citet{di2023handling}\footnote{\url{https://github.com/aledimarco/IXPE-background.git} in order to reduce the instrumental background. We consider a circular region of $60^{''}$ radius centered on the source position to be the `source' region and another concentric annular region between $150^{''}$ and $300^{''}$ to be the background region.} These regions are employed in the \textsc{xpselect} tool of \textsc{ixpeobssim} 31.0.3\footnote{\url{https://ixpeobssim.readthedocs.io/en/latest/}} \citep{baldini2022ixpeobssim} in order to produce cleaned source and background event files for all DUs. We use the latest response file \textsc{ixpe:obssim20240101:v13} for this purpose. In order to extract the polarization properties, different binning algorithms of \textsc{xpbin} are employed. The model-independent PD and PA are estimated using the \textsc{pcube} algorithm \citep{kislat2015analyzing}, while \textsc{xspec} readable count $I$ and Stokes $Q$ and $U$ spectra are generated using the algorithms \textsc{pha1, pha1q} and \textsc{pha1u} respectively. \par 

The $I, Q$ and $U$ spectra are then rebinned using \textsc{ftgrouppha}, where $I$ spectra are first binned using an optimal binning algorithm employing the response matrix, and Stokes spectra are subsequently binned using the $I$ spectrum as a template file.

\begin{deluxetable*}{ccccc}
%\tablenum{1}
\tablecaption{Full observation log for H 1426+428. The columns represent the Instruments used, corresponding Observation IDs, the start and end time for observation (in MJD), and the exposure times.\label{tab:obs}}
\tablewidth{0pt}
\tablehead{
\colhead{Instrument} & \colhead{ObsID} & \colhead{Tstart} & \colhead{Tstop} & \colhead{exposure} \\
\colhead{} & \colhead{} & \colhead{(MJD)} & \colhead{(MJD)} & \colhead{(ks)} \\
}
%\decimalcolnumbers
\startdata
\textit{IXPE} & 03007201 & 60457.33 & 60459.50 & 105.8 \\
\textit{IXPE} & 03007301 & 60496.66 & 60498.83 & 104.4 \\
\textit{Swift}-XRT & 00035020021 & 60457.05 & 60457.66 & 1.6\\ 
\textit{Swift}-XRT & 00030375219 & 	60496.18 & 60496.25 & 1.9\\
\enddata
\end{deluxetable*}

\begin{deluxetable*}{cccccc}
%\tablenum{1}
\tablecaption{The PD and PA for H 1426+428 using all DUs, extracted using the PCUBE algorithm.\label{tab:pcube}}
\tablewidth{0pt}
\tablehead{
\colhead{Obs.} & \nocolhead{} & \colhead{2-8} & \colhead{2-4} &
\colhead{4-6} & \colhead{6-8} \\
\nocolhead{} & \nocolhead{} & \colhead{(keV)} & \colhead{(keV)} &
\colhead{(keV)} & \colhead{(keV)} \\
}
%\decimalcolnumbers
\startdata
03007201 & PD (\%) & $8.7\pm5.2 $ & $3.7\pm4.1 $ & $13.8\pm9.8 $ & $78.9\pm3.0 $ \\
(2024 May 27) & PA (deg) & $21.9\pm16.9$ & $52.4\pm32.7$ & $22.7\pm20.3$ & $7.9\pm11.7$ \\
 & $MDP_{99} (\%)$ & $15.6$  & $14.1$ & $29.7$ & $94.5$ \\
 \hline
03007301 & PD (\%) & $20.3\pm4.1 $  & $18.0\pm3.7 $ & $18.3\pm7.2 $ & $46.1\pm19.5 $ \\
(2024 July 5) & PA (deg) & $-60.6\pm5.7$   & $-61.6\pm5.9$ & $-65.5\pm11.3$ & $-53.1\pm12.3$ \\
 & $MDP_{99} (\%)$ & $12.2$  & $11.1$ & $21.9$ & $59.9$ \\
\enddata
\end{deluxetable*}

\subsection{Swift-XRT}
The \textit{Swift} X-Ray Telescope (XRT) \citep{burrows2005swift} is an X-ray imaging telescope operating at 0.2-10 keV with an effective area of $\sim125~\text{cm}^{-2}$ at 1.5 keV. Contemporaneously with \textit{IXPE}, \textit{Swift}-XRT observed the source for $\sim3.5$ ks. The \textsc{xrtpipeline} task is employed to extract the clean event files, and source and background spectra are generated using \textsc{xselect}. In order to extract the source spectra, we consider a circle of $15^{''}$ radius centered on the source position, and for background, we consider a circular region of radius $40^{''}$ away from the source position. The spectra are rebinned to ensure a minimum of 5 counts in each bin using the \textsc{ftools} task \textsc{grppha}. In order to build the Ancillary Response Files, we have used the task \textsc{xrtmkarf} for every spectral file.   

\section{Results}
We have undertaken both model-independent (\textsc{pcube}) and spectro-polarimetric (using \textsc{xspec}) procedures to explore the polarization features of H 1426+428. In the former model-independent approach, the $I, Q$ and $U$ parameters are computed via summing over the observed photoelectric emission angle for each \textit{IXPE} event, and PD and PA values are subsequently recovered. In \textsc{xspec}, on the other hand, radiative models and polarization models (such as \textsc{polconst}) are employed together to fit the count and Stokes spectra, and from the spectral fit PD and PA values are obtained.

\begin{figure}[ht!]
\begin{center}
  \includegraphics[scale=0.45]{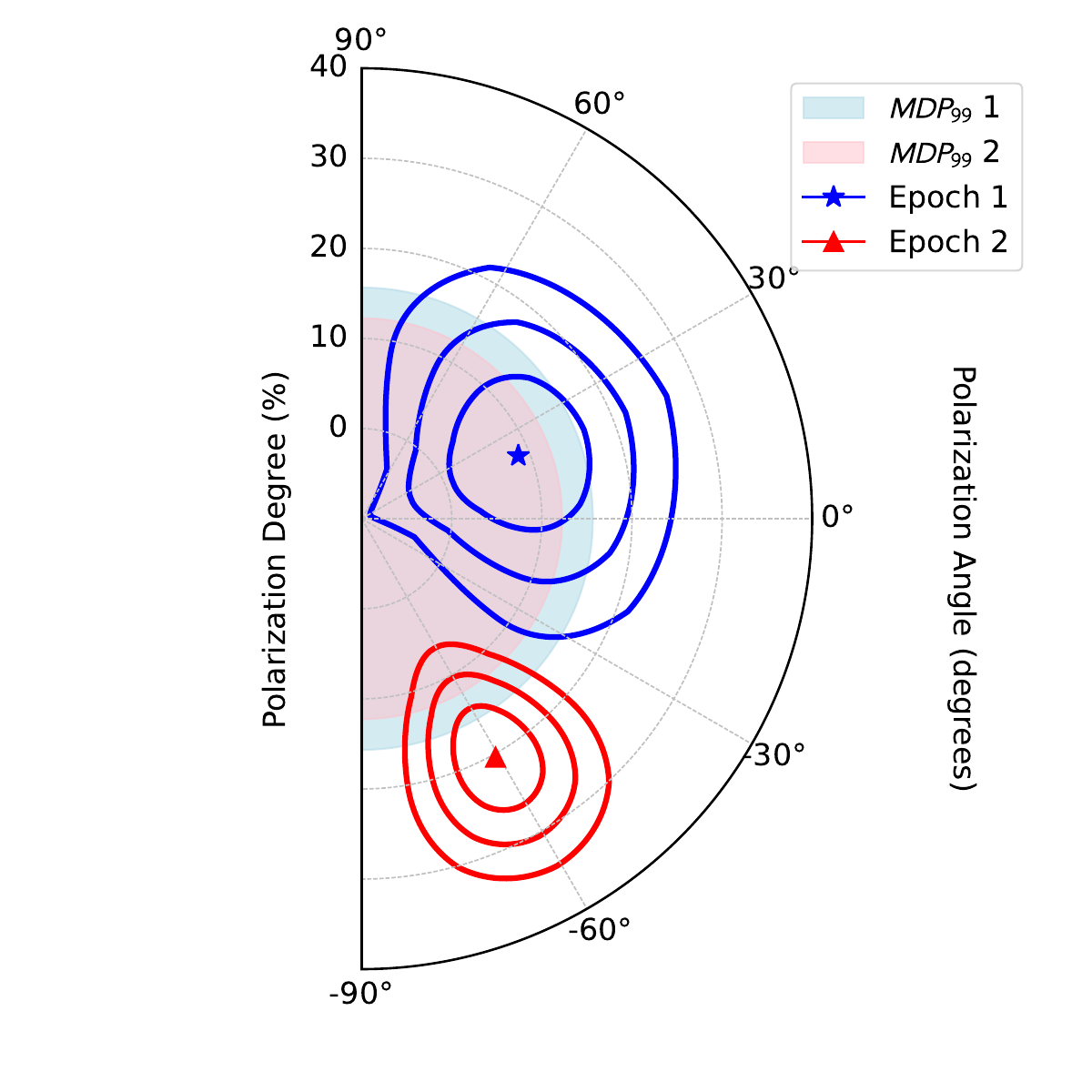}
\caption{Contour plots of PD and PA for H 1426+428 in the 2-8 keV band using \textsc{pcube} algorithm after summing up events from all DUs. The three contours (from inner to outer) represent $1\sigma,2\sigma$ and $3\sigma$ confidence levels. The blue and pink solid semi-circles indicate $\text{MDP}_{99}$ for the first and second observations, respectively. It is apparent that the degree of polarization as obtained from \textsc{pcube} is lower compared to $\text{MDP}_{99}$ during the first measurement but is higher than $\text{MDP}_{99}$ during the second measurement. \label{fig:contours}}  
\end{center}

\end{figure}

\subsection{PCUBE Analysis}
 Our results show that the source is in a relatively higher flux state ($F_{0.3-10\text{keV}} = (5.2\pm0.2)\times10^{-11}~\text{erg cm}^{-2}\text{s}^{-1}$) on 2024 July 05 (O2 hereafter) compared to 2024 May 27 (O1 hereafter), where the flux level was $F_{0.3-10\text{keV}} = (4.1\pm0.3)\times10^{-11}~\text{erg cm}^{-2}\text{s}^{-1}$. As shown in Table \ref{tab:pcube}, the detectable polarization level is different in these two flux states. In the case of O1, in the 2-8 keV band, the PD values are calculated to be $8.7\% \pm 5.2\%$ which are below the threshold given by $\text{MDP}_{99} = 15.6\%$. We, therefore, consider the PD in O1 to be undetected. On the other hand, in the case of O2, PD in 2-8 keV is found to be $20.3\% \pm 4.1\%$, and with $\text{MDP}_{99} = 12.2\%$, we conclude that in the second observation, polarization is significantly measured. There is an indication of a PA shift from O1 ($\sim22^\circ$) to O2 ($\sim-61^\circ$). However, since PD in O1 is undetected, the PA value in O1 cannot be treated as significant. The measurements from all the detectors have been summed up to enhance the confidence interval over PD and PA. In Figure \ref{fig:contours}, we show the $1\sigma, 2\sigma$ and $3\sigma$ confidence contour plots as well as $\text{MDP}_{99}$ for O1 and O2 in the PD-PA plane in 2-8 keV, and corresponding values are listed in Table \ref{tab:pcube}.  Unless explicitly mentioned, the reported error corresponding to each parameter indicates $1\sigma$ confidence level. We have also checked the lightcurves for the presence of variability, but they do no show any significant evidence of intra-observation variability. \par 

To investigate the dependence of PA and PD with energy, we measure the PD and PA values in three energy bands (2-4 keV, 4-6 keV and 6-8 keV) using the \textsc{pcube} algorithm. As shown in Table \ref{tab:pcube}, except for the 2-4 keV in O2, the PD values in all the sub-bands lie consistently below $\text{MDP}_{99}$, therefore preventing us from concluding any energy dependence of PD. 

\begin{deluxetable*}{cccc}
%\tablenum{2}
\tablecaption{ Best fit spectral parameters for \textit{Swift}-XRT and \textit{IXPE} data using \textsc{tbabs*const*polconst*powerlaw} as well as \textsc{tbabs*const*polpow*powerlaw} combination. The errors correspond to $99\%$ confidence on the model parameters.\label{tab:params}}
\tablewidth{0pt}
\tablehead{
\colhead{Component} & \colhead{Parameter} & \colhead{Value} & \colhead{Value}\\
\colhead{} & \colhead{} & \colhead{(2024 May 27)} & \colhead{(2024 July 5)}\\
}
%\decimalcolnumbers
\startdata
%diskbb & $kT_{in}$ & $0.73\pm0.38$ & --   \\
%diskbb & norm & $0.97\pm0.23$ & --   \\
%\hline
power-law & $\Gamma$ & $1.85\pm0.44$ & $1.7\pm0.4$ \\
power law & norm($10^{-3}$) & $17.1\pm1.9$ & $5.0\pm1.0$
\\ 
\hline
polconst & PD$(\%)$ & $2.4\pm1.5$ & $7.0\pm2.1$ \\
polconst & PA$(deg)$ & $39.5\pm33.4$ & $-62.4\pm7.6$ \\
$\chi^2/d.o.f.$ & & 278.4/309 & 294.2/311 \\
\hline
polpow & PD$(\%)$ & $7.6\pm6.2$ & $13.6\pm2.9$ \\
polpow & PA$(deg)$ & $30.7\pm26.1$ & $-62.5\pm7.5$ \\
$\chi^2/d.o.f.$ & & 278.5/309 & 292.6/311 \\
\hline
\enddata
\end{deluxetable*}

\begin{figure}[ht!]
\begin{center}
  \includegraphics[scale=0.45]{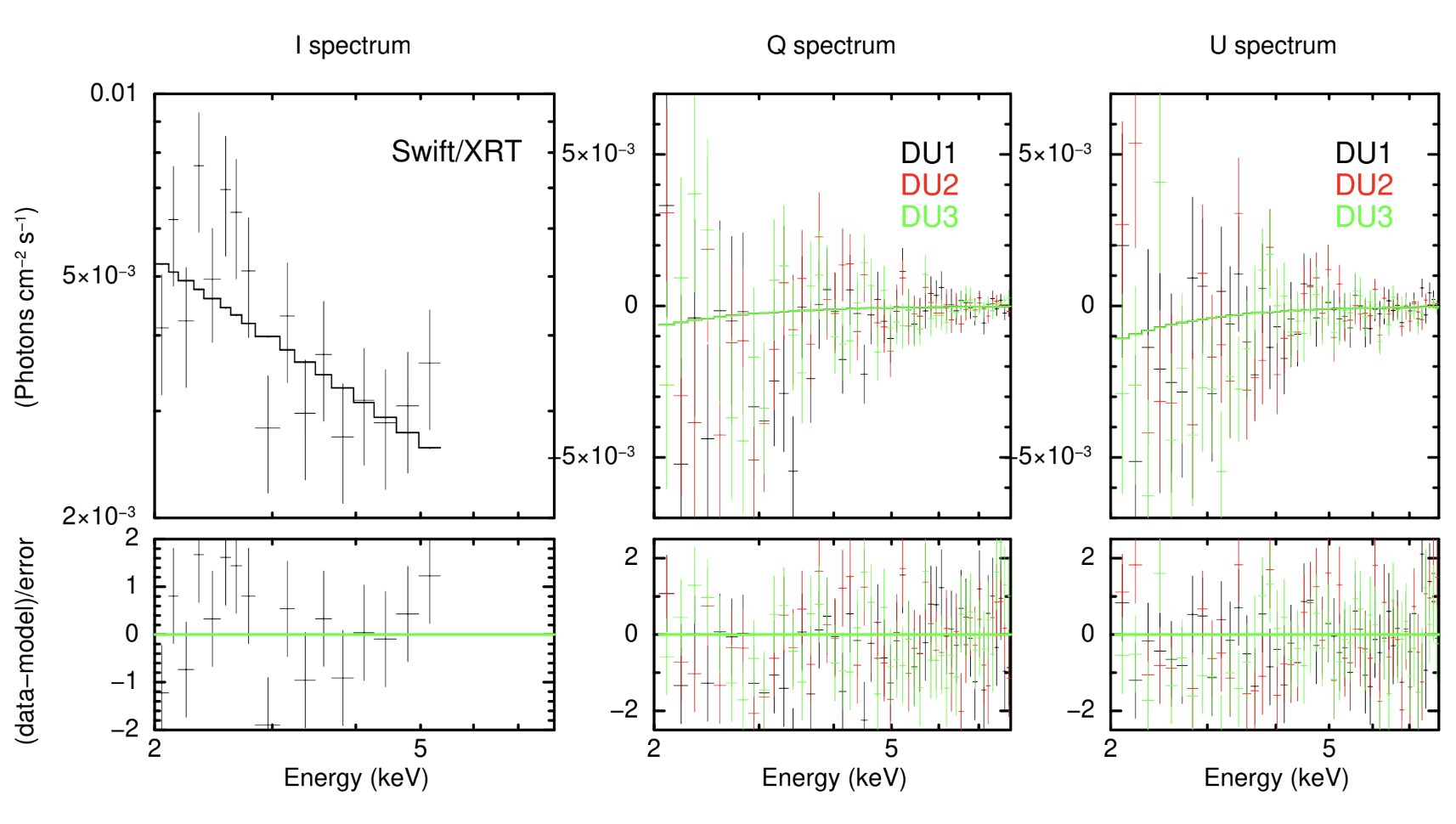}
\caption{ Top left panel shows the \textsc{XSPEC} fit of \textit{Swift}-XRT spectra corresponding to the observation on 2024 May 27. The top middle and top right panels show the simultaneous fit of Stokes $Q$ and $U$ spectra, respectively, including all DUs. Bottom panels show the corresponding fit residuals.\label{fig:xspec}} 

%\caption{Top left panel shows the \textsc{xspec} fit of \textit{Swift}-XRT spectra corresponding to the observation on 2024 May 27. The top middle and top right panels show the simultaneous fit of Stokes $Q$ and $U$ spectra, respectively, including all DUs. Bottom panels show the corresponding fit residuals.\label{fig:xspec}}  
\end{center}

\end{figure}

\subsection{XSPEC Analysis}
In order to explore the basic radiative properties of H 1426+428 during the two observations, we start the spectral fitting of the joint \textit{Swift}-XRT and \textit{IXPE} spectra in the using the model combination \textsc{tbabs*(powerlaw)}, where \textsc{tbabs} is the multiplicative model representing the interstellar absorption, and the \textsc{powerlaw} component represents the multi-color non-thermal Compton emission from the source. For the purpose of our analysis, we fix the interstellar absorption at $9.59\times10^{19}~\text{cm}^{-2}$ (obtained via the HeaSoft hydrogen column density calculator\footnote{\url{https://heasarc.gsfc.nasa.gov/cgibin/Tools/w3nh/w3nh.pl}}). The fitting yields a $\chi^2$ of 563.4 for 309 d.o.f and 658.6 for 311 d.o.f, corresponding to O1 and O2, respectively.\par  %In order to assess the suitability of a more complex physical model like \textsc{nthcomp} to represent the high energy emission, we attempt to fit the data with   \textsc{tbabs*(diskbb+nthcomp)} as well. However, such a model combination provided worse statistics, and the fitted parameters also became unphysical. Therefore, we do not consider such radiative models for spectro-polarimetric analysis. \par 
Subsequently, we fit the $I$ spectra from \textit{Swift}-XRT and $Q, U$ spectra from \textit{IXPE} simultaneously. We attempt the model combination  \textsc{tbabs*const*polconst*(powerlaw)}, where \textsc{polconst} is a multiplicative, energy-independent model for polarization, and \textsc{const} takes care of the cross-calibration factor between \textit{Swift}-XRT and \textit{IXPE} spectra. The fit yielded a $\chi^2$ of 278.4 for 309 d.o.f and 294.2 for 311 d.o.f. corresponding to O1 and O2 respectively. In the case of O1, PD and PA are found to be $2.4\%\pm1.5\%$ and $39.5^\circ\pm33.4^\circ$ respectively. However, since we could not detect PD using the \textsc{pcube} algorithm, the comparison between \textsc{pcube} and \textsc{XSPEC} results could not be made in the case of O1. In O2, the PD value is calculated to be $7.0\%\pm2.1\%$, and PA is calculated to be $-62.4^\circ\pm7.5^\circ$. We see the PD value does not overlap with the corresponding \textsc{pcube} result within the uncertainty. \par

In order to resolve the discrepancy between \textsc{pcube} and \textsc{XSPEC} results, we further test our spectral fitting using the energy dependent polarization (\textsc{polpow}) model. According to the \textsc{polpow} model, $PD(E) = PD_{\rm norm}\times E^{-\alpha_{\rm PD}}$, and  $PA(E) = \psi_{\rm norm}\times E^{-\alpha_{\rm PA}}$, where $PD_{\rm norm}$ is the PD at 1 keV with $\alpha_{\rm PD}$ being the corresponding index. Similarly, $\psi_{\rm norm}$ is the PA at 1 keV with $\alpha_{\rm PA}$ being the corresponding index. Since we do not observe energy dependence of PA, we pivot the index of PA to zero. This yields the best-fit spectra with $\chi^2/dof = 278.5/309, A_{\rm PD} = 0.02 \pm 0.01, \alpha_{\rm PD} = -0.14 \pm 0.07$ for O1, and $\chi^2/dof = 292.6/311, A_{\rm PD} = 0.04 \pm 0.01, \alpha_{\rm PD} = -0.39 \pm 0.8$ for O2. As previously mentioned, comparison between \textsc{pcube} and \textsc{xspec} results could not be undertaken for O1 due to the non-detection of polarization. For O2, the model is integrated in 2-8 keV with the best-fit parameters to obtain the PD $\sim13\%$ and PA $\sim-62^\circ$. This matches with the \textsc{pcube} result within the uncertainty limit. \par

In Table \ref{tab:params}, we provide the best fit parameters alongside the respective errors obtained using \textit{XSPEC} fitting. The simultaneously fitted $I, Q$ and $U$ spectra for all the DUs as well as the residuals pertaining to the observation on 2024 May 27 are shown in Figure \ref{fig:xspec}. 

\begin{deluxetable*}{cccc}
%\tablenum{2}
\tablecaption{Details of optical polarimetric measurements. Pertaining to each observation, we provide the magnitude, degree and angle of polarization.\label{tab:optical}}
\tablewidth{0pt}
\tablehead{
\colhead{Obs. date} & \colhead{Magnitude} & \colhead{PD (\%)} & \colhead{PA (deg)}\\
}
%\decimalcolnumbers
\startdata
2024 May 27 & $15.91\pm0.04$ & $1.56\pm0.27$ & $38.6\pm4.6$   \\
2024 May 28 & $15.90\pm0.01$ & $1.49\pm0.13$ & $48.3\pm5.4$   \\
2024 July 5 & $15.95\pm0.02$ & $2.47\pm0.06$ & $74.1\pm5.0$   \\
2024 July 6 & $15.95\pm0.02$ & $1.87\pm0.07$ & $75.6\pm4.4$   \\
\enddata
\end{deluxetable*}

\subsection{Results from Optical Polarimetry}
During the \textit{IXPE} pointings, we coordinated an optical campaign with Observatorio de Sierra Nevada (OSN), Calar Alto Observatory and the 1.8 m Perkins Telescope to obtain contemporaneous polarimetric data in the optical domain. Photometric observations from OSN were obtained in the R-band using the DIPOL-1 instrument installed at the 0.9 m T90 telescope and analyzed using standard procedures \citep{otero2024characterization}. The photometric BVRI data from OSN were taken with the 1.5 m T150 telescope We also obtain data from the Calar Alto Observatory (CAHA) captured with the CAFOS instrument installed on the 2.2 m telescope, which allows for photo-polarimetric observations. The polarimetry was performed in the R band, with additional photometric BVRI observations \citep{pedrosa2024iop4}. R-band polarimetric observations as well as BVRI photometry were carried out using the PRISM camera mounted on the Perkins Telescope Observatory employing the polarimetric analysis procedures as described in \citet{jorstad2010flaring} and differential photometry with the comparison stars 1, 2, and 3 from \citet{smith1991ubvri} to estimate polarization parameters and brightness. During both O1 and O2, the R-band magnitude lies within $\sim$15-16, which agrees with the previously observed magnitude for the target \citep{remillard1989two}. During both O1 and O2, the R-band polarization degree is found to be much lower ($<3\%$) compared to the X-ray polarization degree. Also, the PA in optical during O2 {\rm differs} significantly from the PA in X-ray. The details of optical magnitude, polarization degree and angle during the observation campaign are provided in Table \ref{tab:optical}.

\section{Discussion and Summary}
We have presented the first \textit{IXPE} observations of the TeV detected EHSP blazar H 1426+428
obtained in 2024. All of our observations were complemented with contemporaneous optical and \textit{Swift}/XRT campaigns. During the $\sim$40 day difference between the two observations, the source demonstrated an enhancement in X-ray flux. During our first observation on 2024 May 27, the polarization degree and angle remained undetected. However, during the second \textit{IXPE} pointing on 2024 July 5, significant ($>20\%$) polarization in X-ray could be detected with $>99\%$ confidence. The X-ray PD values were found to be significantly higher compared to those in the optical. \par 

The higher X-ray PD measurements compared to those in other wavelengths are common among \textit{IXPE}-observed HSP blazars, and have been observed in several cases, e.g. Mrk 421 \citep{di2022x}, Mrk 501 \citep{chen2024x}, 1ES 0229+20 \citep{ehlert2023x} and 1ES 1959+65 \citep{errando2024detection}. The increasing degree of polarization from optical to X-ray is generally interpreted in terms of acceleration of high energy electrons across  shock. Higher energy photons emitted in X-ray are generated by synchrotron emission from electrons in the immediate downstream of the shock front, while lower energy photons such as optical, are generated by electrons further downstream because of the longer radiative cooling time, where they are subjected to more turbulent component of the magnetic field \citep{marscher1985models,marscher2021frequency}. Within the X-ray domain, we attempt to decipher any plausible dependence of PD on energy. During both of our observations, the energy-resolved PD values lie below $\text{MDP}_{99}$, and consequently, we can not make any definitive claim of energy dependence. In the optical domain, since we measure the polarization value only in $R$-band, we are unable to infer any energy dependence of polarization. We also note that the detected polarization ($\sim20\%$) in X-ray is far lower than the synchrotron limit. With a power-law energy spectrum of relativistic electron of the form $N(E) \propto E^{-p}$ with $p$ being the spectral index, the expected degree of polarization is $\sim\frac{p+1}{p+7/3} \sim 70\%$ for $p=2$ and $\sim75\%$ for $p=3$, if the relativistic electrons are in a site of homogenous magnetic field \citep{rybicki1991radiative}. Within a region of the jet where the fraction of the ordered component of the magnetic field is $f_{\text{ord}}$, the mean degree of polarization is given by:
\begin{equation}
  \langle P \rangle \propto [f_{\text{ord}}^2 + (1-f_{\text{ord}})^2N^{-1}(\nu)]^{1/2},
\end{equation}
where the number of turbulent cells involved in the emission at frequency $\nu$ is $N(\nu) \sim \nu^{-1/2}$ \citep{marscher2022linear}. In case the disordered component of the magnetic field dominates the ordered component, we can expect $P \sim \nu^{1/4}$. Therefore, we would expect a higher degree of polarization in the X-ray domain compared to optical. In the X-ray band, we have $\nu \sim 10^{18}$, while in optical R-band, $\nu \sim 4.6\times10^{14}$. If we consider the ratio, then $P_X/P_O \sim (ratio)^{1/4} \sim 6.8$. Given we are getting $\sim 2\%$ polarization in optical, $P_X \sim 13.6\%$ matches up with the $P_X$ value as determined from the spectral fit (within uncertainties). This implies that the X-ray emission originates from a partially ordered (by shock compression) region close to the shock acceleration site, while the optical emission arises from a more strongly turbulent region farther from the shock, which reduces the degree of polarization \citep{marscher2021frequency,marscher2022linear}. Recent high-resolution 3D RHD simulations of relativistic jets also show such shock mediated onset of turbulence, and a corresponding rise in turbulent component of estimated magnetic field \citep{ayan2024}. \par 

In our analysis, we observe a change in the polarization angle in the optical domain, while in the X-ray domain, we measure the polarization angle only during O2. Substantial rotation of polarization angle in X-ray has previously been detected only in the case of Mrk 421 so far \citep{di2023discovery}, although a number of optical rotations have been reported in the literature \citep{blinov2016robopol}. One specific model to explain such a feature is the propagation of a magnetosonic shock along the direction of a helical magnetic field in the jet, where the rate of PA rotation is determined by the time taken by the shock to cross one full pitch \citep{zhang2016polarization}. Another interesting scenario supporting PA rotation could be the compression of the magnetic field structure in the post-shock region, where the {\bf ordered component of the field} dominates the turbulent component \citep{angelakis2016robopol}. However, in order to confirm such a scenario, simultaneous VLBA observation of PA in the radio domain is necessary. Previous VLBA campaign on TeV blazars during the years 2001-2004, a faint jet extending to the northwest, with a position angle of about $-25^\circ$ was obtained for our source of interest at 8.4 GHz \citep{piner2008parsec}, which is roughly orthogonal to the PA in the optical domain as per our observation. In previous IXPE observation of HSP blazars, PA in optical roughly aligns with the jet position angle \citep{liodakis2022polarized,abe2024insights}, although large rotations in PA have also been observed \citep{di2023discovery,middei2023ixpe}. Alignment of optical PA with the jet position angle in the case of BL Lac objects is generally interpreted in the shock-compressed scenario where the ordered magnetic field is aligned perpendicular to the jet axis \citep{marscher2021frequency}. Misalignment between the jet direction and PA is generally attributed to plausible local disturbances \citep{piner2005vlba}. However, we now have growing evidence that the position angle of radio jets can change over time, e.g. OJ 287 \citep{britzen2018oj287}, PG 1553+113 \citep{lico2020parsec} and so on \citep{weaver2022kinematics}, in the absence of the simultaneous VLBA data, we are unable to confirm if the position angle is along the projected jet
direction, or if PA is off-axis relative to the jet. Models of turbulent plasma undergoing a shock transition can explain the alignment of PA with the jet axis \citep{marscher2014turbulent,tavecchio2021probing}. Further multi-wavelength observation with simultaneous VLBA campaigns is therefore crucial to confirm or reject such a hypothesis. \par 

The significantly higher polarization in the X-ray domain compared to the optical obtained in our study, therefore, strongly supports the energy-stratified shock acceleration scenario in the HSP blazar jets. Observation of rotation of the polarization angle indicates the plausibility of the propagation of shocked plasma along a section of the jet with a helical magnetic field.

\section{Acknowledgement}

 We thank the anonymous referee for constructive suggestions that have improved the quality of the manuscript. AB and MA acknowledge funding under NASA contract 80NSSC24K1745. J.O.S. acknowledges financial support from the projetct ref. AST22\_00001\_9 with founding from the European Union - NextGenerationEU, the \textit{Ministerio de Ciencia, Innovación y Universidades, Plan de Recuperaci\'{o}n, Transformación y Resiliencia}, the \textit{Consejer\'{i}a de Universidad, Investigación e Innovación} from the \textit{Junta de Andalucía} and the \textit{Consejo Superior de Investigaciones Científicas}; as well as from INFN Cap. U.1.01.01.01.009. A.D. is thankful for the
support of the Proyecto PID2021- 433126536OA-I00 funded by MCIN / AEI /10.13039/501100011033. The IAA-CSIC co-authors acknowledge financial support from MCIN/AEI/ 10.13039/501100011033 through the Center of Excellence Severo Ochoa award for the Instituto de Astrof\'{i}isica de Andaluc\'{i}a-CSIC (CEX2021-001131-S), and through grants PID2019-107847RB-C44 and PID2022-139117NB-C44. Some of the data are based on observations collected at the Observatorio de Sierra Nevada, which is owned and operated by the Instituto de Astrof\'isica de Andaluc\'ia (IAA-CSIC). S.M. research activity is carried out with contribution of the Next Generation EU funds within the National Recovery and Resilience Plan (PNRR), Mission 4 - Education and Research, Component 2 - From Research to Business (M4C2), Investment Line 3.1 - Strengthening and creation of Research Infrastructures, Project IR0000012 – ``CTA+ - Cherenkov Telescope Array Plus".
Based on observations collected at Centro Astron\'omico Hispano en Andalucía (CAHA) at Calar Alto, proposals 24A-2.2-011 and 24B-2.2-015, operated jointly by Junta de Andaluc\'ia and Consejo Superior de Investigaciones Cient\'ificas (IAA-CSIC). The research at Boston University was supported in part by the National Science Foundation grant AST-2108622. This study was based in part on observations conducted using the 1.8m Perkins Telescope Observatory (PTO) in Arizona, which is owned and operated by Boston University. The scientific results reported in this article are based on observations made by the X-ray observatories \textit{Swift}-XRT and \textit{IXPE}. We acknowledge the use of the software packages DS9 and HEASoft. 

\bibliography{sample631}{}
\bibliographystyle{aasjournal}

\end{document}